\documentclass{article}

\usepackage{arxiv}

\usepackage[utf8]{inputenc} % allow utf-8 input
\usepackage[T1]{fontenc}    % use 8-bit T1 fonts
\usepackage[colorlinks,allcolors=black,pdftex]{hyperref}       % hyperlinks
\usepackage{url}            % simple URL typesetting
\usepackage{booktabs}       % professional-quality tables
\usepackage{amsfonts}       % blackboard math symbols
\usepackage{nicefrac}       % compact symbols for 1/2, etc.
\usepackage{microtype}      % microtypography
\usepackage{lipsum}

\usepackage{graphicx}%
\usepackage{multirow}%
\usepackage{amsmath,amssymb,amsfonts}%
\usepackage{amsthm}%
\usepackage{mathrsfs}%

\graphicspath{ {./FIGs/} }

\usepackage{algorithm}%
\usepackage{algorithmicx}%
\usepackage{algpseudocode}%
\usepackage{listings}%
\usepackage{caption}
\usepackage{subcaption}

\newcommand{\refsub}[2]{\ref{#1}#2}

\title{Automatic extraction of wall streamlines from oil-flow visualizations using a convolutional neural network}

\author{
 Jonas Schulte-Sasse\\
  Institute of Aeronautics and Astronautics\\
  Technical University Berlin\\
  Straße des 17. Juni 135, 10623 Berlin \\
  \texttt{jonasschs.15@campus.tu-berlin.de} \\
   \And
 Ben Steinfurth\\
  Institute of Aeronautics and Astronautics\\
  Technical University Berlin\\
  Straße des 17. Juni 135, 10623 Berlin \\
  \texttt{ben.steinfurth@tu-berlin.de} \\
   \And
 Julien Weiss\\
  Institute of Aeronautics and Astronautics\\
  Technical University Berlin\\
  Straße des 17. Juni 135, 10623 Berlin \\
  \texttt{julien.weiss@tu-berlin.de} \\
}

\begin{document}
\maketitle
\begin{abstract}
Oil-flow visualizations represent a simple means to reveal time-averaged wall streamline patterns. Yet, the evaluation of such images can be a time-consuming process and is subjective to human perception. In this article, we present a fast and robust method to obtain quantitative insight based on qualitative oil-flow visualizations. Using a convolutional neural network, the local flow direction is predicted based on the oil-flow texture. This was achieved with supervised training based on an extensive dataset involving approximately one million image patches that cover variations of the flow direction, the wall shear-stress magnitude and the oil-flow mixture. For a test dataset that is distinct from the training data, the mean prediction error of the flow direction is as low as three degrees. A reliable performance is also noted when the model is applied to oil-flow visualizations obtained from the literature, demonstrating the generalizability required for an application in diverse flow configurations.
\end{abstract}

% keywords can be removed
\keywords{Wall shear-stress measurement \and Flow visualization \and Artificial neural networks}

\section{Introduction}\label{sec:intro}
The wall shear-stress plays a pivotal role in governing the interaction between a fluid and its solid boundary, influencing flow patterns, material wear, and surface integrity. In cardiovascular physiology, for instance, the wall shear-stress is a determinant in vascular health and disease, as abnormal shear forces contribute to endothelial dysfunction, arterial remodeling, and the development of atherosclerosis \cite{Papaioannou.2005}. Similarly, in industrial applications, such as pipeline transport, understanding the influence of operating conditions on the wall shear-stress is essential for optimizing flow efficiency and preventing erosion or structural failure.

In view of its ubiquitous presence across various disciplines, the experimental quantification of the wall shear-stress remains a complex challenge in experimental fluid mechanics \cite{Groe.2008,Zhao.2019,Orlu.2020,Rudolph.2009}. The task becomes even more difficult when the distribution over a solid surface is required to reveal the footprints of intricate flow structures rather than single-point measurements. For this purpose, the oil-flow method represents an attractive alternative \cite{Lu.2010} to quantitative measurements. Here, a thin layer of oil, typically mixed with colored particles, is applied to the surface of interest. The shear force exerted by the flow then causes the oil layer to displace along the direction of wall streamlines. As the oil spreads, streaks are formed revealing the surface flow pattern and thereby allowing for qualitative insight into the wall shear-stress distribution. Frequently used in wind tunnel testing, oil-flow visualizations have been performed to study the near-wall flow over airfoils \cite{Selig.2012}, turbine blades \cite{Hecklau.2012} and separating flows \cite{Simmons.2022, Steinfurth.2024}, among many other examples.

Contrasting with the relative simplicity in execution, the post-processing of oil-flow visualization images can be an elaborate process. Typically, lines are drawn manually on the oil-flow image such as to capture a surface flow pattern that is consistent with the visual perception. While humans are very apt pattern recognizers, this manual approach remains inherently subjective and may be the source of misleading interpretations.

Alternative to the classical, analog evaluation is the utilization of digital processing tools. For instance, quantitative data can be obtained from flow visualizations with tufts by adopting an edge detection scheme \cite{Steinfurth.2020} or Deep Learning (DL) techniques \cite{Tsalicoglou.2022}. Relying on  artificial neural networks to represent complex functions, a tremendous potential for advancing experimental techniques can be ascribed to the latter \cite{Vinuesa.2023}. As for oil-flow visualizations, two digital processing techniques have been proposed. First, wall streamlines can be reconstructed by computing the displacement field of particles supplied to the oil suspension based on cross-correlation \cite{Mosharov.2005}. Second, optical flow techniques were adopted recently to extract quantitative information from oil-flow images by evaluating the temporal development of luminescent intensity \cite{Rohlfs.2024}. The shared drawback of these two techniques is that they require time sequence data showing the motion of particles contained in the oil layer. Furthermore, the acquisition needs to be performed while the oil-flow pattern is still forming, and recording parameters need to be selected carefully to capture the motion of particles.

In the present study, we strive to establish a novel technique for the prediction of wall streamlines from oil-flow visualizations only requiring single-image input. Specifically, a convolutional neural network (CNN), representing a class that is typically used for computer vision tasks \cite{LeCun.2015}, is employed to learn the relationship between oil-flow texture and flow direction.

The structure of this article is as follows. First, in section \ref{sec:methods}, we explain the CNN set-up, training scheme and choice of hyperparameters. Then, the performance of the novel technique is examined by evaluating oil-flow images obtained in our lab as well as images taken from published articles (section \ref{sec:results}). Concluding remarks are provided in section \ref{sec:disscusion}.
\section{Set-up and training of neural network}\label{sec:methods}
In this section, the function principle of the method proposed in this article is explained. Furthermore, we elaborate on the conditions defining the network training.

\subsection{Function principle}
\begin{figure}[H]
    \centering
    \includegraphics[width=17cm,keepaspectratio]{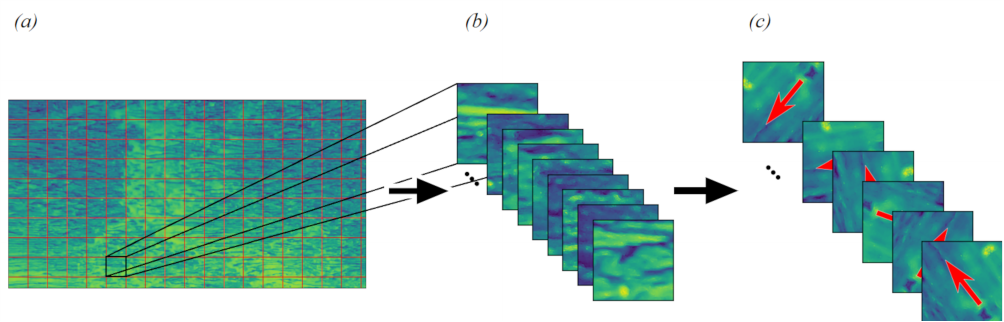}
    \caption{Schematical representation of function principle (a) oil-flow image sub-divided into square patches, (b) patches are supplied as an input to the CNN, (c) for each patch, the flow direction is predicted (highlighted by red arrows)}\label{fig:DataPre}
\end{figure}

In the context of the present study, a CNN learns the relationship between an oil-flow pattern and the flow direction which may be defined by

\begin{equation}
    \Psi _\vartheta (\Omega ) = \varphi
\end{equation}

where $\Psi$ is the function represented by the network subject to its weights and biases $\vartheta$, $\Omega$ is a two-dimensional oil-flow visualization patch and $\varphi$ is the angle defining the  streamline direction. To obtain the dense wall streamline distribution from a single oil-flow visualization image, it needs to be subdivided into several patches as shown in Fig.~\ref{fig:DataPre}. Subsequent to the subdivision, each patch is supplied individually to the model that assigns the flow direction. The choice of appropriate patch dimensions will be discussed in detail later on in section~\ref{sec:preprocessing}.

\begin{figure}[H]
    \centering
    \includegraphics[width=17cm,keepaspectratio]{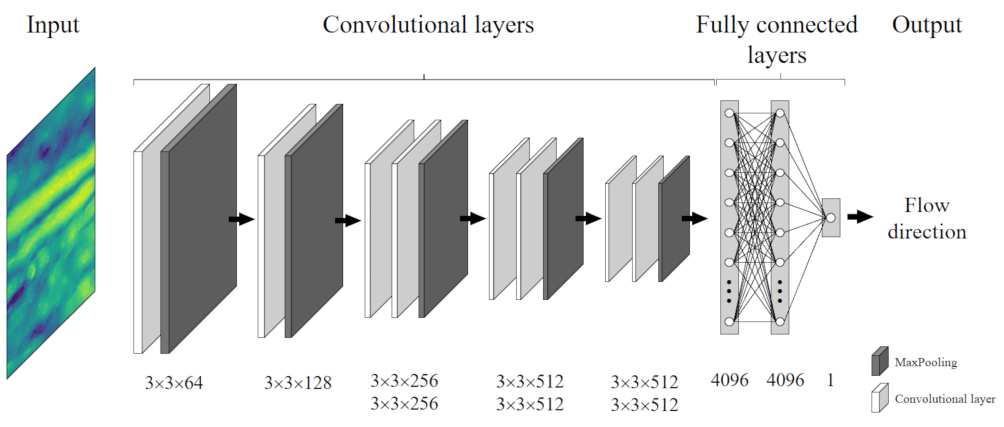}
    \caption{Model architecture; numbers of kernels and dimensions stated at the bottom of the figure, example: $3\times\ 3\times 64$ stands for 64 filters with dimensions of $3\times 3$}\label{fig:VGG11}
\end{figure}

Showing some analogies with the visual perception of animals \cite{Hubel.1959}, CNNs are inherently suited to grasp the relevant information contained in multi-dimensional fields. The network architecture employed in this study is shown in Figure~\ref{fig:VGG11}. It is derived from the standard Visual Geometry Group (VGG) set-up, specifically the VGG-A variant \cite{Simonyan.04.09.2014} being one of the most popular image recognition networks. 

The input is provided in the shape of two-dimensional oil-flow patches of dimensions $184\times 184\, \mathrm{px^2}$. Note that the \textit{viridis} colormap (blue-green-yellow) is used throughout this article to highlight oil-flow images/patches although they only contain one color channel.
The input patches are supplied to the CNN, consisting of eight convolutional layers, in part followed by max pooling layers, and three fully-connected layers. 

\begin{enumerate}
    \item The convolutional layers are the building blocks responsible for extracting features from the input data, which is achieved by $3\times 3$ filters sliding over the input. At each position, the filter weights are multiplied by the corresponding input entries before the products are summed. This process helps detect patterns like edges or other simple objects.
    \item Max pooling layers, on the other hand, are used to downsample the output of convolutional layers. Here, the max pooling operation selects the maximum value inside $2\times 2$ windows, thereby reducing the dimension while retaining the most important information.
    \item After passing the final max pooling layer, the output is flattened into a one-dimensional vector and passed through two fully-connected layers where the output of every neuron from the first layer is transmitted to every neuron from the second layer. For each neuron, one specific weight and bias are applied, followed by a ReLU activation function.
\end{enumerate}

The final layer consists of only one linearly activated neuron whose output is interpreted as the flow angle (in degrees). To ensure a reliable prediction, the network is trained in a supervised fashion as explained in the next section.

\subsection{Network training}
Labeled data are required for the CNN to learn the mapping between oil-flow patches and flow direction. To this end, reference visualization images are recorded in a blow-down wind tunnel designed for the calibration of wall shear-stress sensors \cite{Weiss.2024}. The rectangular channel flow facility brings the advantage of a homogeneous wall shear-stress field that is known in terms of direction and magnitude. In the wind tunnel, the wall shear-stress magnitude is easily varied to expand the variance pertaining to the dataset and making it applicable to different flow conditions. For the same reason, various compositions of the oil-flow suspension are used by changing the ratio between oil, thinner and UV-active color particles. The suspension was applied to a flat surface of dimensions $220\times 130\,\mathrm{mm^2}$, and images were taken for 18 configurations with a camera whose optical axis was oriented normal with respect to the test section surface. The $5200\times3000\, \mathrm{px^2}$ images were then divided into approximately 500 patches of dimensions $184\times 184\,\mathrm{px^2}$. Since the camera angle was not changed for the 18 configurations, only images with a single flow direction (e.g., from left to right) were recorded. To incorporate training data for different flow directions, each patch was rotated digitally in the range $\varphi = 0^\circ - 357^\circ$ with an increment of $\Delta \varphi = 3^\circ$, yielding a dataset of approximately one million patches, which was split into 85\% training, 10\% validation and 5\% test datasets. The training data were used to adjust the weights and biases while the validation dataset ensured that the CNN would not overfit to the training data. The test data were held out to examine the performance (section~\ref{sec:test}).

\begin{figure}
    \centering
    \includegraphics[width=8.5cm,keepaspectratio]{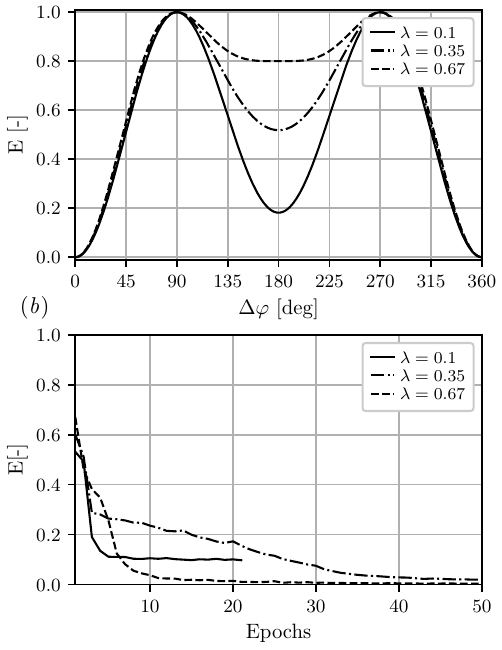}
    \caption{(a) Loss values depending on hyperparameter lambda, (b) learning progress for different loss functions}\label{fig:lossfun}
\end{figure}

The definition of the loss function is a key aspect to the model’s accuracy since this error is considered to iteratively update the network \cite{Rumelhart.1986}. In the present study, the loss function was designed to exhibit penalization peaks at multiples of $90^\circ$ where the predicted flow angle and the true angle are perpendicular. In contrast, the error at multiples of $180^\circ$ (correct prediction of direction but incorrect sense of direction) should be penalized to a lesser extent. We observed that an alternative loss function where the error is simply proportional to the deviation in flow angles (i.e., loss maxima at multiples of $180^\circ$) would not ensure a stable training process. Furthermore, the loss function was required to be bounded, periodic, rotation-invariant and differentiable. Satisfying these criteria, we chose the function

\begin{eqnarray*}
     E(\Delta\varphi) = \sin^2{\Delta\varphi} + \lambda\cos^3{\Delta\varphi}
\end{eqnarray*}

where $\Delta \varphi$ is the deviation between predicted and true flow angle, and the coefficient $\lambda$ is used to modulate the local minima at multiples of $\Delta\varphi=180^\circ$. The effect of three values of $\lambda$ on the loss function is illustrated in Figure~\ref{fig:lossfun}{a}. 

Applying the loss functions with $\lambda = (0.1, 0.35, 0.67)$ presented in Figure~\ref{fig:lossfun}{a}, the network was trained by providing mini-batches of 128 patches and by minimizing the loss using Adam optimization at a learning rate of $2\cdot 10^{-4}$. The training was terminated by adopting the \textit{early stopping} technique such as to avoid overfitting. Specifically, the network was not updated further when the validation data loss did not decrease for ten subsequent epochs, indicating conditions where the model does not generalize to unseen data. The development of the mean loss per epoch is shown in Figure~\ref{fig:lossfun}{b}. Clearly, the most accurate predictions, with a mean absolute deviation below three degrees, is achieved for $\lambda = 0.67$. We therefore proceeded with this value, which is also the largest value that preserves local minima at multiples of $180^\circ$ deviation.

\begin{figure}
    \centering
    \includegraphics[width=17cm,keepaspectratio]{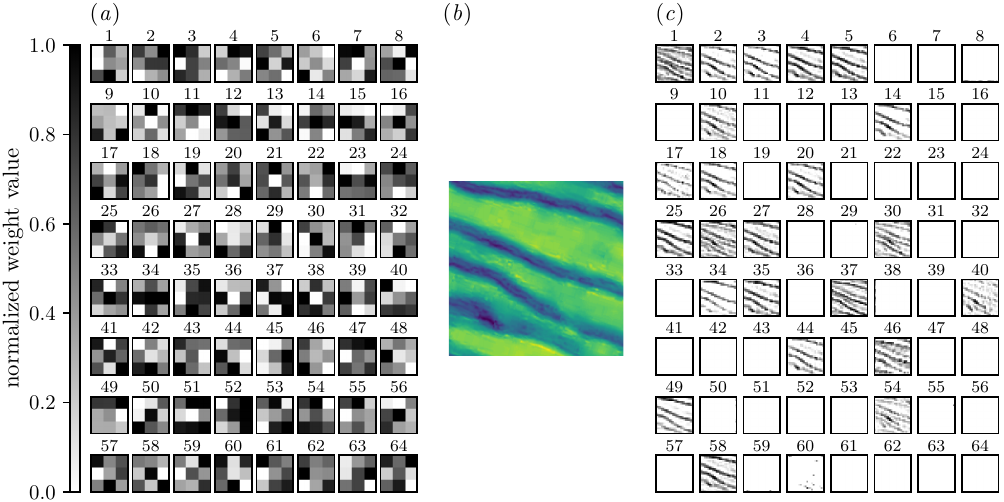}
    \caption{Illustration of feature extraction: (a) kernels of first convolutional layer, (b) example oil-flow patch, (c) feature maps correponding to kernels}\label{fig:featuremap}
\end{figure}

Figure~\ref{fig:featuremap} visualizes the feature extraction for the initial convolutional layer of the trained network where the most elementary shapes present in the input images are extracted.

% Fig_4_alt.pdf: feature map (first 64 fitler) after 3rd convolutional block (256)
% Fig_4.pdf: feature map (first 64 fitler) after 1st convolutional block (256)
  The normalized weights pertaining to the 64 kernels are shown in the left of the figure, and the corresponding feature maps for an example patch are shown in the right. While some convolutional filters appear to extract no tangible information for this particular patch (e.g., filters 2 and 3), others clearly capture the streaks contained in the oil-flow pattern (e.g., filters 21 and 22). This information is further transmitted through the subsequent layers and eventually associated with the local flow direction.

\subsection{Hyperparameters}\label{subsec:spatial}
In the following, we report relevant hyperparameters of the proposed model, including the network dimensions.

The main parameters defining the network training are stated in Table~\ref{tab:constParamters}.
 
\begin{table}
\caption{Training hyperparameters}
    \centering
    \begin{tabular}{c|c|c|c}
       Batch size & Learning rate & Learning scheduler & Scheduler momentum\\\hline
        128  &        0.0002 &        exponential &               0.96\\
    \end{tabular}
    \label{tab:constParamters}
\end{table}

\begin{figure}
    \centering
    \includegraphics[width=8.5cm,keepaspectratio]{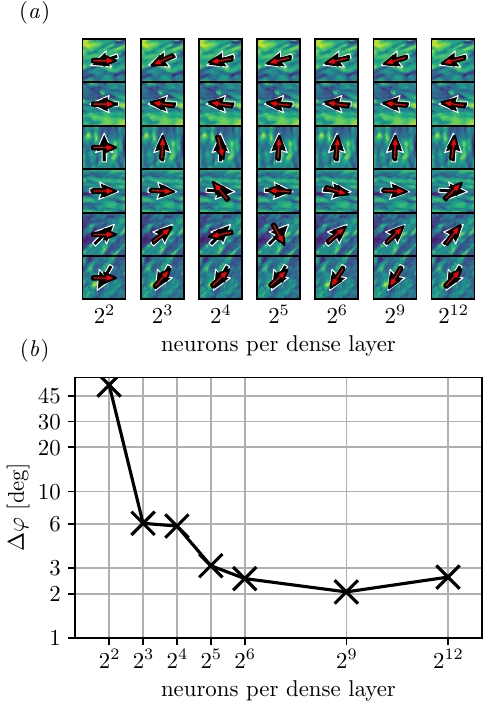}
    \caption{Effect of number of neurons in fully-connected layers; (a) predicted (red) over true (black) for different numbers of neurons in fully-connected layers, (b) mean deviation between predicted and true flow direction at the end of training for varied network width }\label{fig:width}
\end{figure}

The network's width determines the capacity to learn complex flow structures. Strongly affecting the computational cost associated with the model inference, we investigated the influence of the number of neurons in the first two fully-connected layers. Note that the standard VGG architecture has $2^{12}$ units per layer. Figure~\ref{fig:width} reveals that the number of neurons can be lowered significantly as mean absolute deviations below three degrees are also reached with only $2^5$ neurons per layer. Only for even smaller numbers of neurons do we note a performance degradation as the narrowest network with as few as four units per layer yields a deviation on the order of $45^\circ$, which is illustrated in Figure~\refsub{fig:width}{a} indicating the predicted (red) along with the true (black) flow direction for selected sample patches.

Based on the test results summarized in Figure~\ref{fig:width}{b}, the value of $2^9$ neurons per fully-connected layer was selected for the model as it yields the highest accuracy and results in a lighter network compared to the standard VGG architecture.

In contrast to the number of neurons in the fully-connected layers, the influence of the number of kernels on the computational cost is negligible and did not affect the model accuracy significantly.

\begin{figure}
    \centering
    \includegraphics[width=8.5cm,keepaspectratio]{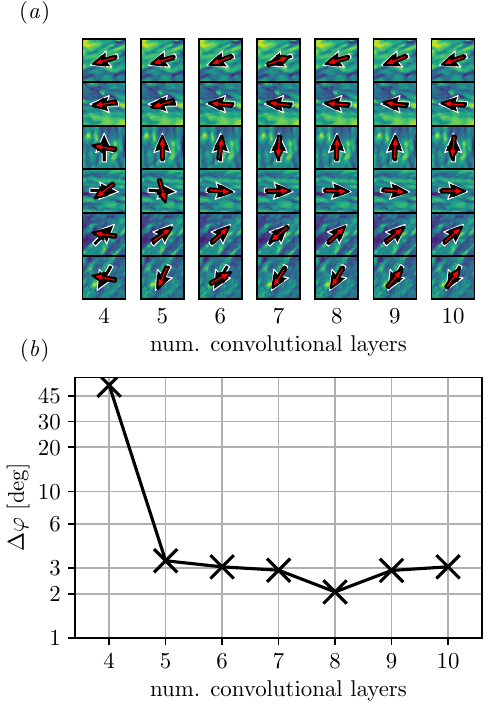}
    \caption{(a) predicted (red) over true (black) flow orientation for varied number of convolutional layers, (b) prediction error at the end of training for varied network depth}\label{fig:depth}
\end{figure}

Finally, the network depth, represented by the number of convolutional layers, was adjusted (Figure~\ref{fig:depth}). While doing so, we retain the principle of combining either one or two convolutional layers with one max pooling operation.

Interestingly, neither adding nor subtracting two convolutional layers affects the accuracy of the model significantly compared to the standard VGG-A architecture with eight layers. However, a shallow network with only four convolutional layers is not capable of capturing and extracting the flow patterns accurately. On the other hand, a deeper network with 10 layers does not yield a higher accuracy. Therefore, we proceeded with eight layers as shown in Figure~\ref{fig:VGG11}.

\subsection{Image pre-processing}\label{sec:preprocessing}
The flow direction can only be predicted accurately when sufficient information is provided to the model. Specifically, visualization patches need to be in focus, and the illumination and choice of oil-flow suspension must be adjusted such as to ensure sufficient contrast. In addition to these standard guidelines that also apply to classical qualitative oil-flow visualizations, the physical patch dimensions need to be considered for the method presented in this article. While smaller patches justify the linearization of wall streamlines that is inherently conducted with this method, it also yields a higher spatial resolution. On the other hand, patch dimensions need to be chosen sufficiently large to contain sufficient information.

\begin{figure}
    \centering
     \includegraphics[width=8.5cm,keepaspectratio]{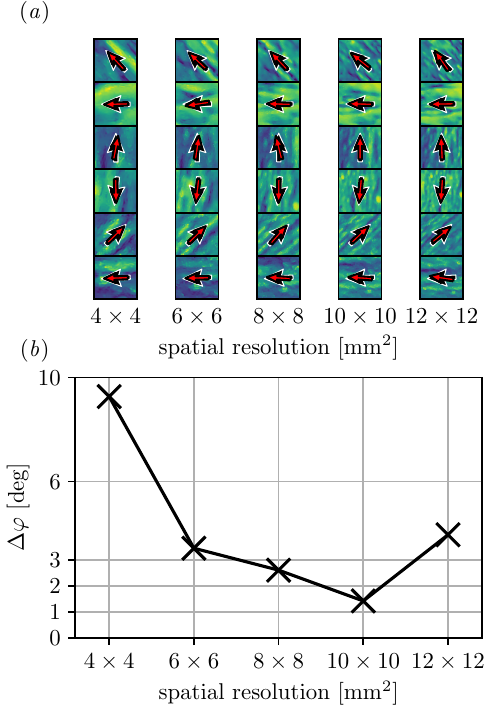}
    \caption{(a) predicted (red) over true (black) flow orientation for selection of patch sizes, (b) prediction error at the end of training depending on patch dimensions}\label{fig:spatial}
\end{figure}

To establish some recommendations for suitable physical patch dimensions, we systematically assessed the model output for different patch sizes, ranging from $4\times 4\,\mathrm{mm^2}$ to $12\times 12\,\mathrm{mm^2}$ (Figure~\ref{fig:spatial}).

While the smallest investigated patch size of $4 \times 4 \,\mathrm{mm^2}$ results in significant deviations of almost 10 degrees, mean absolute deviations on an order of three degrees are observed for all the larger dimensions. It appears that, at least for the training dataset, edge lengths of $6\,\mathrm{mm}$ ensure that sufficient information regarding the flow pattern is provided.

It is important to note that the model only accepts input patches of $184 \times 184 \,\mathrm{px^2}$. Therefore, depending on the optical magnification of the recording set-up, raw images may have to be be interpolated or sub-sampled.

\subsection{Data postprocessing}
\label{sec:postprocessing}

Based on the results for the training dataset presented above, a reasonable accuracy can be attested. However, to ensure a homogenized output prediction, a universal outlier scheme may be applied to the raw output data. The classical algorithm used in this study is presented in Ref. \cite{Westerweel.2005}, evaluating neighbouring data points to detect outliers. Three parameters need to be specified: $\gamma$ as a threshold deviation with respect to neighbors, above which the point is marked as an outlier, $\epsilon$ as noise estimate and $r$ as search radius. The results presented in the following section were achieved with $\gamma = 1$, $\epsilon = 0.5$ and $r=2$.

The algorithm is an iterative process that executes until all points become valid according to the specified threshold. Initially, points are marked as outliers. Then, for each outlier, the number of valid neighbors within the specified radius is calculated. All outliers are sorted in such a way that the number of outlying neighbors increases. Data points with more than 50\% valid neighbors are then substituted in this order by the median of valid neighbors, ensuring that information flows from regions with presumed high accuracy to regions with lower accuracy. In the subsequent loop, if the total number does not change, the threshold $\gamma$ is slightly modified, allowing the algorithm to converge to a solution.
\section{Performance assessment}\label{sec:results}
In the following, we estimate the performance of our method. First, the flow direction is predicted for patches of the test dataset where the true angle is known. While the test patches are distinct from the training and validation data, a general similarity can still be expected since they were acquired with the same experimental set-up. Therefore, we also test whether the method can make reliable predictions for other experimental conditions by considering the separating flow over two different backward-facing ramps \cite{Steinfurth.2024, Simmons.2022} and a delta wing \cite{Kumar.1121202111232021}.

\subsection{Test dataset}\label{sec:test}
\begin{figure}[H]
    \centering
    \includegraphics[width=8.5cm,keepaspectratio]{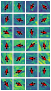}
    \caption{Prediction (red) over true (black) flow direction for selected test patches}
    \label{fig:testset}
\end{figure}

Recall that the test dataset consists of ca. 50,000 patches neither used to train nor to validate the model. Figure~\ref{fig:testset} shows 28 selected test patches with true (black) and predicted (red) directions. The average absolute deviation for the patches shown in this figure is $\Delta \varphi \approx4.3^\circ$. For the entire test dataset, the mean absolute deviation is $\Delta \varphi \approx2.7^\circ$, indicating excellent prediction capabilities when the oil-flow images are similar in recording settings to the training data. It is perhaps surprising to note that the model accurately predicts the sense of flow direction despite only evaluating single-image input data. This unexpected but welcome effect is probably caused by marginal characteristics that are not visible by the human eye.

\subsection{Generalizability} \label{subsec:generalization}
Generalizability is an important concept in DL that refers to the ability of a model to make accurate predictions when the input differs from the training data.

\begin{figure}[H]
    \centering
    \includegraphics[width=8.5cm,keepaspectratio]{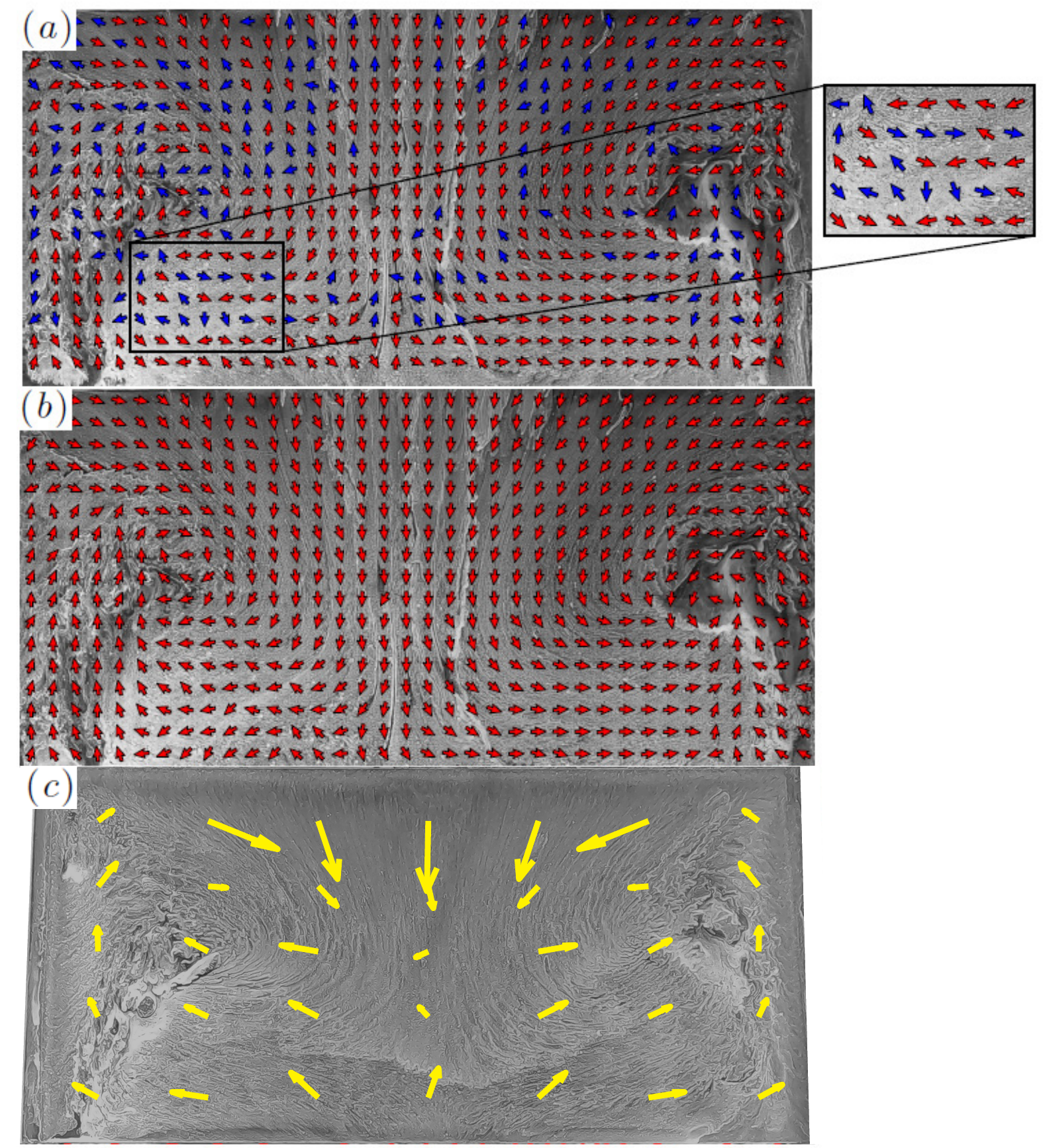}
    \caption{Wall streamline extraction for a backward-facing ramp set-up, (a) valid model output shown in red, outliers shown in blue (b) vector field with outliers substituted (c) previous oil-flow visualization overlayed with wall shear-stress measurement data, each vector represents one sensor location}\label{fig:test20}
\end{figure}

The oil-flow visualization of the surface flow in a one-sided diffuser is the first test case. Despite the two-dimensionality of the test section, the flow is known to be highly three-dimensional due to the effects of side-walls \cite{Steinfurth.2024}. Specifically, flow reversal near the side walls reaches further upstream than in the symmetry plane of the test section, giving rise to large-scale footprints of secondary flow that are observed in the oil-flow visualization (Figure~\ref{fig:test20}). Input patches have dimensions of $10 \times 10\,\mathrm{mm^2}$ (i.e., they are similar in size as the training patches).

While delivering a reasonable representation of the wall streamline pattern, there is a proportion of about 25\% predictions that are classified as outliers (highlighted by blue arrows in Figure \ref{fig:test20}(a)). After substituting the outliers (Figure \ref{fig:test20}(b)), a streamline pattern is obtained that is in good agreement with wall shear-stress measurements presented in Figure \ref{fig:test20}(c). This comparison also demonstrates that the method proposed in this article yields a much denser representation of the wall streamline field than elaborate measurements.

\begin{figure}[H]
    \centering
    \includegraphics[width=8.5cm,keepaspectratio]{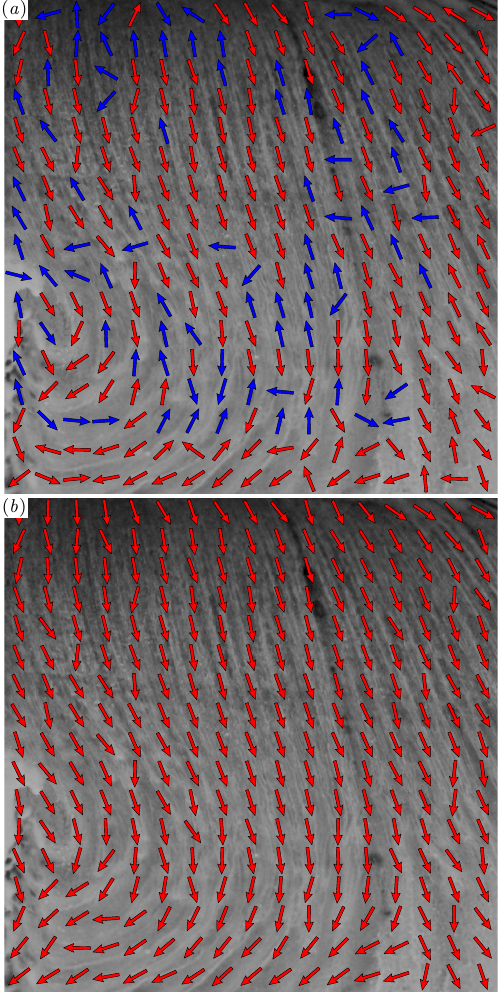}
    \caption{Flow over a curve ramp taken from \cite{Simmons.2022}, air flow from top to bottom. (a) valid model output shown in red, outliers shown in blue (b) prediction field with outliers substituted}\label{fig:testnasa}
\end{figure}

For the second generalizability test case, we consider the oil-flow visualizations not produced in our lab but instead taken from a published article \cite{Simmons.2022} addressing the flow past a backward-facing ramp (Figure~\ref{fig:testnasa}). Here, the turbulent boundary layer experiences a deflection and undergoes separation, forming a recirculation zone on the left-hand side of the image. 

The oil-flow texture in this test case appears to be more blurred than in the training data. Yet, the raw prediction (Figure\refsub{fig:testnasa}{a}) only contains a minor proportion of outliers that are highlighted in blue. Near the side-wall located on the left-hand side of the image, the flow reversal is not indicated accurately and the outlier detection does not yield satisfying results due to the small dimensions of this flow structure. Nevertheless, a satisfying representation of the forward-directed flow spanning the majority of this visualization is achieved.

\begin{figure}[H]
    \centering
    \includegraphics[width=8.5cm,keepaspectratio]{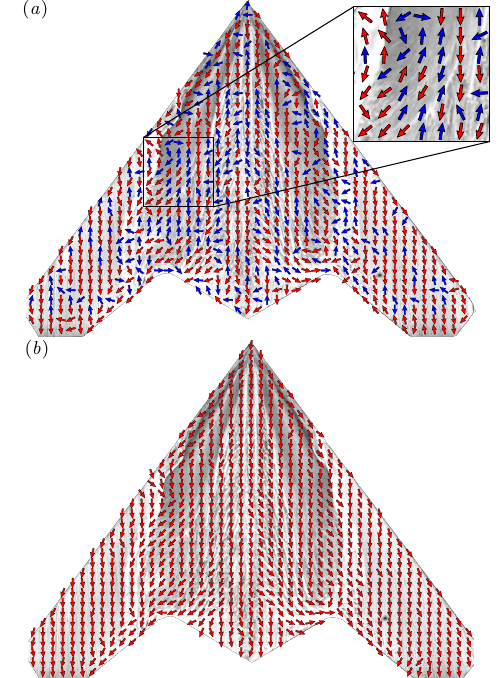}
    \caption{Oil-flow visualization over a non-slender flying wing, flow from top to bottom \cite{Kumar.1121202111232021}. (a) valid model output shown in red, outliers shown in blue (b) vector field with outliers substituted}\label{fig:d23}
\end{figure}

The final test case was also obtained from a published image and involves a lambda wing at $23^\circ$ angle of attack for $Re=0.5\cdot10^6$. The occurring primary vortex, the flow separation regions on the trailing edge of the main wing-body and the wake-like surface structure on the side wingtips are investigated in Ref.~\cite{Kumar.1121202111232021}.  In figure \refsub{fig:d23}{a}, the prediction of the surface flow pattern for this model is shown. Compared to the other test cases presented so far, a larger proportion of outliers is present, which can be explained by the lack of information contained in the oil-flow visualization inside large regions on the wing surface (see detail figure). However, the outlier detection algorithm succeeds in delivering a plausible flow pattern because there is a sufficient amount of reliable data points in the initial model output.

\section{Conclusions and outlook}\label{sec:disscusion}

The objective in this study was to develop a DL-based technique that allows to extract quantitative information, in the shape of the local flow direction, from qualitative oil-flow visualizations. To this end, a CNN was trained with an extensive database recorded during reference experiments where the flow direction was set.

Both for the training and test data, the mean absolute deviation between the flow direction predicted by the model and the true direction was below three degrees. Perhaps most surprisingly, the network is able to not only determine the flow direction but also its sense of direction. Considering the single-image nature of the analysis, the same is not necessarily expected in human-level analysis.

When diverse visualization images, acquired under experimental conditions distinct from the present study, are considered, a larger proportion of false predictions is observed. This is especially the case when the oil-flow texture becomes ambiguous, also affecting the qualitative analysis of such images. This effect can be alleviated by classical outlier detection \cite{Westerweel.2005} and substitution as long as physically implausible predictions do not dominate.

The main advantage of the new technique is that it allows for an automated processing of either new oil-flow experiments or past images that may have been acquired many years ago.

\bibliographystyle{IEEEtran}  
\bibliography{sn-bibliography}  %%% Remove comment to use the external .bib file (using bibtex).
%%% and comment out the ``thebibliography'' section.

\end{document}